\documentclass{mem}
\usepackage{natbib}\usepackage{txfonts}\usepackage{balance}
\usepackage{graphicx}
\usepackage[a4paper,breaklinks,dvipdfm]{hyperref}
\idline{75}{282}
\begin{document}
\def\teff{$T\rm_{eff }$}
\def\kms{$\mathrm {km s}^{-1}$}

\title{
Cloud structure of brown dwarfs from spectroscopic variability observations\thanks{Based on observations made with the NASA/ESA Hubble Space Telescope, obtained at the Space Telescope Science Institute, which is operated by the Association of Universities for Research in Astronomy, Inc., under NASA contract NAS 5-26555. These observations are associated with program \# 12314 and 13280}
}

%   \subtitle{}

\author{
E.\,Buenzli\inst{1} 
\and M.~S.\,Marley\inst{2}
\and D.\,Apai\inst{3}
\and R.~E.\,Lupu\inst{4}
} 

\institute{
Max Planck Institute for Astronomy,
Koenigstuhl 17,
69117 Heidelberg,
Germany,
\email{buenzli@mpia.de}
\and
NASA Ames Research Center,
Moffet Field, CA 94035,
USA 
\and
University of Arizona,
Tucson, AZ 85721,
USA 
\and
SETI Institute / NASA Ames Research Center,
Moffet Field, CA 94035,
USA 
}

\authorrunning{Buenzli et al. }

\titlerunning{Cloud structure of brown dwarfs}

\abstract{
Recent discoveries of variable brown dwarfs have provided us with a new window into their three-dimensional cloud structure. The highest variables are found at the L/T transition, where the cloud cover is thought to break up, but variability has been found to occur also for both cloudy L dwarfs and (mostly) cloud-free mid T dwarfs. We summarize results from recent HST programs measuring the spectral variability of brown dwarfs in the near-infrared and compare to results from ground-based programs. We discuss the patchy cloud structure of L/T transition objects, for which it is becoming increasingly certain that the variability does not arise from cloud holes into the deep hot regions but from varying cloud thickness. We present a new patchy cloud model to explain the spectral variability of 2MASSJ21392676+0220226. We also discuss the curious multi-wavelength variability behavior of the recently discovered very nearby early T dwarf WISE J104915.57-531906.1B (Luhman 16B) and the mid T dwarf 2MASS J22282889-431026.
}
\maketitle{}

\section{Introduction}

The discovery of brown dwarfs with significant near-infrared variability \citep[e.g.][]{artigau09, radigan12} has indicated that clouds are not homogeneously distributed in some brown dwarf atmospheres. While L dwarf photospheres are thought to be covered by thick silicate and iron clouds, these clouds have disappeared by mid-T spectral types. One explanation to simultaneously explain the color evolution through the L/T transition, near-infrared variability and the re-emergence of the 0.99 $\mu$m FeH band in early to mid T dwarfs is the growth of holes in the clouds that allow flux from deeper, hotter regions to emerge \citep{ackerman01, burgasser02, marley10}. Early T dwarfs indeed seem to be the most frequent strong variables \citep{radigan14}, but significant variability has also been found in cloudy L and (mostly) clear mid-T dwarfs \citep{buenzli12, heinze13, buenzli14, radigan14, wilson14}. Furthermore, spectral variability measured with HST \citep{buenzli12, apai13} is inconsistent with deep cloud holes. 

\section{Are L/T transition objects partly cloudy?}

The first two confident detections of variable brown dwarfs interpreted as patchy cloud coverage were the T2.5 dwarf SIMP0136 \citep{artigau09} and the T1.5 dwarf 2M2139 \citep{radigan12}, both lying squarely in the L/T transition. For SIMP0136, the amplitude of the variations in J and K band (peak-to-peak $\approx$6\% and $3\%$) could be explained by several model combinations \citep{radigan12} with combinations of cloudy and clear models, as well as with different sedimentation efficiency parameters \citep[$f_{sed}$,][]{ackerman01}, corresponding to different cloud thickness. For 2M2139, much larger amplitude variations in J, H and K band (between $\approx 15-30\%$ depending on wavelength and observing date) combined with a spectrum suggested that atmospheres with fully clear sections could not well reproduce the observations \citep{radigan12}. Many combinations of cloud thicknesses and temperatures remained possible.

Spectral variability observations obtained with HST/WFC3 of both 2M2139 and SIMP0136 \citep{apai13} from about 1.08 to 1.66~$\mu$m revealed that the variability amplitude is significantly lower in the deep water absorption band at 1.4~$\mu$m than in J or H band, but otherwise remarkably constant outside of the water absorption feature. For both objects the characteristics are very similar except for the amplitude, and we only focus on 2M2139 (Fig. 1). The water band centered at 1.15 $\mu$m varies only very slightly less than the J band peak emission, and no difference at all is seen in the K I feature at 1.25 $\mu$m compared to the continuum. The ratio is only marginally smaller in the H band than the J band peak. 

\begin{figure*}[t!]
\resizebox{\hsize}{!}{\includegraphics[clip=true]{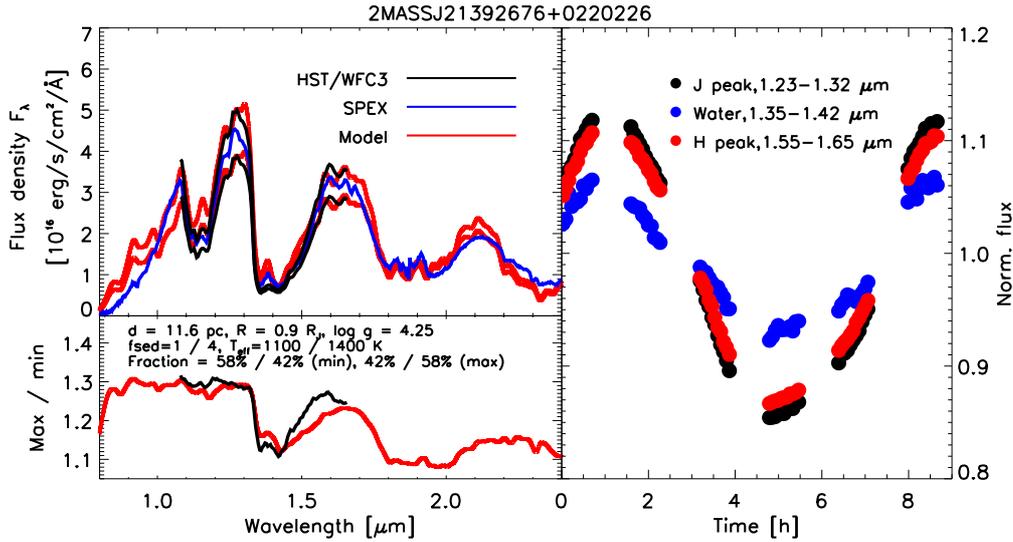}}
\caption{\footnotesize
Top left: Maximum and minimum HST/WFC3 spectrum (black) for 2MASSJ21392676+0220226 and an average spectrum from the SPEX prism library \citep[blue,][]{burgasser06}. Overplotted is our best patchy cloud model (red). Bottom left: Ratio of maximum to minimum spectrum for the observations (black) and the model (red) with parameters indicated. Right: Measured light curve over 6 HST orbits integrated over three wavelength regions.}
\label{fig1}
\end{figure*}

\citet{apai13} also showed that any combination of cloudy and clear models could be excluded. Combinations of thin (E-type, $T_{\mathrm{eff}}=1100$~K) and thick (B-type, $T_{\mathrm{eff}}=800$~K) clouds  \citep{burrows06} could well reproduce the color variations. However, these models could not fit the variability in absorption bands, and the very thick and cold B-type cloud is rather atypical for brown dwarfs. Here we model for the first time the full spectral variability. We find that a model combination with approximately equal covering fraction of a cool thick cloud ($f_{\mathrm{sed}} = 1,\,T_{\mathrm{eff}} = 1100$) and a thin warmer cloud ($f_{\mathrm{sed}} = 4,\, T_{\mathrm{eff}}=1400$ K) provides the best match to both the spectrum and the variability (Fig. \ref{fig1}). A sole increase in the fractional coverage of the thin cloud cannot explain the further evolution through the L/T transition. At a slightly later stage, formation of deeper holes or additional thinning of the cloud is still needed to explain the bluer color and a re-emergence of FeH. 

Our model manages to explain most of the characteristics of the spectral variability, while other combinations of $f_{\mathrm{sed}}$ fail to match the variability ratio across the J band and/or the relative amplitude between the 1.4~$\mu$m band and the J band. No model adequately reproduces the variability on the red side of the 1.4 $\mu$m water feature, nor the spectral shape at $1.3-1.5$~$\mu$m and $1.13-1.2$~$\mu$m, perhaps due to incomplete model opacities. The predicted K-band amplitude is consistent with observations \citep{radigan12}, but the spectral mis-match in K band suggests that models including vertical mixing \citep{stephens09} might improve the fit. Furthermore, these model combinations are not self-consistent in the sense that the two models have different, independent temperature-pressure profiles. An attempt at more self-consistent patchy cloud models was made in \citet{marley10}, but only for partly cloudy atmospheres with fully clear holes. These models cannot adequately reproduce our observed spectral variability. Calculations of self-consistent patchy cloud models with different cloud thicknesses should be the next step in the modeling of brown dwarf spectral variability.

\subsection{The curious case of Luhman 16B}

The recent detection of the very nearby binary L/T transition dwarf WISEJ1049 \citep{luhman13}, aka Luhman 16AB, has provided an extraordinary benchmark object for the detailed study of cloud structure at the L/T transition. The T0.5 type B component \citep{burgasser13} was found variable in i+z band \citep{gillon13} with very fast light curve evolution. Simultaneous multi-wavelength photometry with the GROND instrument \citep{biller13} revealed a behavior not in line with the spectroscopic observations of the previously discussed two variable brown dwarfs with similar spectral type. Particularly curious is a non-detection of variability in the J band on one night (but strong detection one week earlier), while significant variability was simultaneously found in z' and H together with anti-correlated variability in r' and i' and out-of-phase variability in K band. No current patchy cloud model can produce significant variability ($\approx$10\%) in z' and H band but no or significantly lower variability in J band. The out-of-phase variability in K band is also different than for 2M2139, where quasi-simultaneous observations by \citet{radigan12} suggest that JHK light curves are all in phase. No comparable observations at r' and i' exist for the other two brown dwarfs because they are too faint, and thus anti-correlation at these wavelengths may not be unusual. \citet{biller13} propose that the phase shift correlates with the probed pressure, similar to the T6.5 dwarf 2M2228 (cf Sect. \ref{t6}). However, preliminary analysis of new HST spectroscopic variability observations of Luhman 16B (Buenzli et al. in prep) show that the variability in the water band at 1.4~$\mu$m, which probes pressures even lower than the r' and i' bands, is completely in phase with the J and H band variability. In fact, the HST spectral variability appears to be remarkably similar for Luhman 16B to 2M2139 (Fig. \ref{fig1}) and SIMP0136, contrary to the GROND observations. Furthermore, \citep{burgasser14} obtained a 45 min ground-based spectroscopic variability sequence that also suggested the strongest variability at Y and J band with some decrease towards H and K band. If the GROND observations are correct, Luhman 16B undergoes drastic changes not only in the light curve shape, but also in its spectral variability characteristics that currently cannot be explained by patchy cloud models. 

\section{Variability beyond the L/T transition}
\label{t6}
Silicate clouds are thought to have sunk below the visible photosphere beyond spectral type of $\approx$T4, but substantial variability has been observed in several such brown dwarfs. The most notable is the T6.5 dwarf 2M2228, discovered as variable in J band by \citet{clarke08} and characterized in detail with simultaneous HST spectral and Spitzer photometric time series by \citet{buenzli12}. Some of its variability may be explained by patchy sulfide clouds that can potentially condense at these temperatures \citep{morley12}. However, the largest variability amplitude is found in the water band at 1.4 $\mu$m, anti-correlated to the variability in the  J and H band peak. Furthermore, the light curves in the IRAC 4.5 $\mu$m channel and in the methane absorption band at 1.65~$\mu$m have intermediate phases between J, H and the water band, indicating a correlation of the shift with the probed atmospheric pressure. 

Simple patchy cloud models are unlikely to be able to match these observations, and more complex circulation patterns that invoke temperature perturbations may also play a role. \citet{robinson14} investigated if a periodic temperature perturbation as predicted by dynamical models \citep{showman13} could propagate upward and introduce phase dependent variability. While their model currently neglects clouds and rotation, it shows that variability with the approximate amplitude and phase shift can, in principle, arise in this manner. Additional mid T dwarfs have been found variable in J band \citep{radigan14} or in H$_2$O or CH$_4$ bands \citep{buenzli14}, suggesting that 2M2228 is not unique. 

\section{Conclusions}

Brown dwarf variability is ubiquitous and points to complex cloud structure and evolution. Most current cloud and circulation models are one-dimensional and cannot yet sufficiently model these effects, but significant progress is already underway. The discovery of the unusually bright and variable Luhman 16B has provided a target that allows a very detailed view into the cloud structure at the L/T transition. It is one of very few L/T transition dwarfs accessible to Gaia, which will provide important points for obtaining the binary orbit. This will eventually lead to an independent mass measurement, crucial for calibrating models. 
 
\begin{acknowledgements}
EB acknowledges support from the Swiss National Science Foundation (SNSF). Support for HST program \#12314 and 13280 was provided by NASA through a grant from the Space Telescope Science Institute, which is operated by the Association of Universities for Research in Astronomy, Incorporated, under NASA contract NAS5-26555. This research has benefitted from the SpeX Prism Spectral Libraries, maintained by Adam Burgasser at http://pono.ucsd.edu/\~{}adam/browndwarfs/spexprism.
\end{acknowledgements}

\bibliographystyle{aa}

\begin{thebibliography}{}

\bibitem[Ackerman 
\& Marley(2001)]{ackerman01} Ackerman, A.~S., \& Marley, M.~S.\ 2001, \apj, 556, 872 

\bibitem[Apai et al.(2013)]{apai13} Apai, D.,  et al.\ 2013, \apj, 768, 121 

\bibitem[Artigau et al.(2009)]{artigau09} Artigau, {\'E}., 
et al.\ 2009, \apj, 701, 1534 

\bibitem[Biller et al.(2013)]{biller13} Biller, B.~A., 
 et al.\ 2013, \apjl, 778, L10 

\bibitem[Buenzli et al.(2012)]{buenzli12} Buenzli, E., et al.\ 2012, \apjl, 760, L31 

\bibitem[Buenzli et al.(2014)]{buenzli14} Buenzli, E., et al.\ 2014, \apj, 782, 77 

\bibitem[Burgasser et al.(2014)]{burgasser14} Burgasser, A.~J., 
et al.\ 2014, \apj, 785, 48 

\bibitem[Burgasser et al.(2006)]{burgasser06} Burgasser, A.~J., 
et al.\ 2006, \apj, 637, 1067 

\bibitem[Burgasser et al.(2002)]{burgasser02} Burgasser, A.~J., 
 et al.\ 2002, \apjl, 571, L151 

\bibitem[Burgasser et al.(2013)]{burgasser13} Burgasser, A.~J., 
Sheppard, S.~S., \& Luhman, K.~L.\ 2013, \apj, 772, 129 

\bibitem[Burrows et al.(2006)]{burrows06} Burrows, A., Sudarsky, 
D., \& Hubeny, I.\ 2006, \apj, 640, 1063 

\bibitem[Clarke et al.(2008)]{clarke08} Clarke, F.~J., et al.\ 2008, \mnras, 386, 2009 

\bibitem[Gillon et al.(2013)]{gillon13} Gillon, M., et al.\ 2013, \aap, 555, L5 

\bibitem[Heinze et al.(2013)]{heinze13} Heinze, A.~N., et al.\ 2013, \apj, 767, 173 

\bibitem[Luhman(2013)]{luhman13} Luhman, K.~L.\ 2013, \apjl, 
767, L1 

\bibitem[Marley et al.(2010)]{marley10} Marley, M.~S., Saumon, 
D., \& Goldblatt, C.\ 2010, \apjl, 723, L117 

\bibitem[Morley et al.(2012)]{morley12} Morley, C.~V., et al.\ 2012, \apj, 756, 172 

\bibitem[Radigan et al.(2012)]{radigan12} Radigan, J., 
et al.\ 2012, \apj, 750, 105 

\bibitem[Radigan et al.(2014)]{radigan14} Radigan, J., 
et al.\ 2014, \apj \, in press 

\bibitem[Robinson 
\& Marley(2014)]{robinson14} Robinson, T.~D., \& Marley, M.~S.\ 2014, \apj, 785, 158 

\bibitem[Showman 
\& Kaspi(2013)]{showman13} Showman, A.~P., \& Kaspi, Y.\ 2013, \apj, 776, 85 

\bibitem[Stephens et al.(2009)]{stephens09} Stephens, D.~C., 
et al.\ 2009, \apj, 702, 154 

\bibitem[Wilson et al.(2014)]{wilson14} Wilson, P.~A., Rajan, 
A., \& Patience, J.\ 2014, A\&A in press


\end{thebibliography}

\end{document}